\documentclass[prb,amsmath,amssymb,superscriptaddress,twocolumn]{revtex4}
\usepackage{graphicx}

\newcommand{\br}{{\bf r}}
\newcommand{\bx}{{\bf x}}

\newcommand{\bk}{{\bf k}}

\newcommand{\bp}{{\bf p}}

\newcommand{\bq}{{\bf q}}

\newcommand{\bK}{{\bf K}}

\DeclareMathAlphabet{\mathpzc}{OT1}{pzc}{m}{it} 
\begin{document}
\title{Many-body instability of Coulomb interacting bilayer graphene: RG approach}
\author{Oskar Vafek}
\author{Kun Yang}
\affiliation{National High Magnetic Field Laboratory and Department
of Physics, Florida State University, Tallahassee, Florida 32306,
USA}
\address{}

\date{\today}
\begin{abstract}
Low-energy electronic structure of (unbiased) bilayer graphene is
made of two Fermi points with {\em quadratic} dispersions, if
trigonal-warping and other high order contributions are ignored. We
show that as a result of this qualitative difference from
single-layer graphene, short-range (or screened Coulomb)
interactions are marginally {\em relevant}. We use renormalization
group to study their effects on low-energy properties of the system,
and show that the two quadratic Fermi points spontaneously split
into four Dirac points, at zero temperature. This results in a
nematic state that spontaneously breaks the six-fold lattice
rotation symmetry (combined with layer permutation) down to a
two-fold one, with a finite transition temperature. Critical
properties of the transition and effects of trigonal warping are
also discussed.
\end{abstract}

 \maketitle
The ability to predict the nature of the low temperature state of an
interacting quantum system is one of the main goals of condensed
matter theory. Nevertheless, despite ongoing effort, no single
method has proved universally sufficient and experimental input is
essentially inevitable.

Under special circumstances, however, progress can be made. In
particular, in non-interacting systems with susceptibilities
diverging as the temperature approaches zero, the inclusion of
arbitrarily small interaction can be shown to lead to a finite, but
also arbitrarily small transition temperature. The method of choice
in this case is the renormalization group (RG), which has the virtue
of unbiased determination of the leading
instability\cite{Shankar.RevModPhys.66.129.1994}.

\begin{figure}[t]
\begin{center}
\begin{tabular}{cc}
\includegraphics[width=0.19\textwidth]{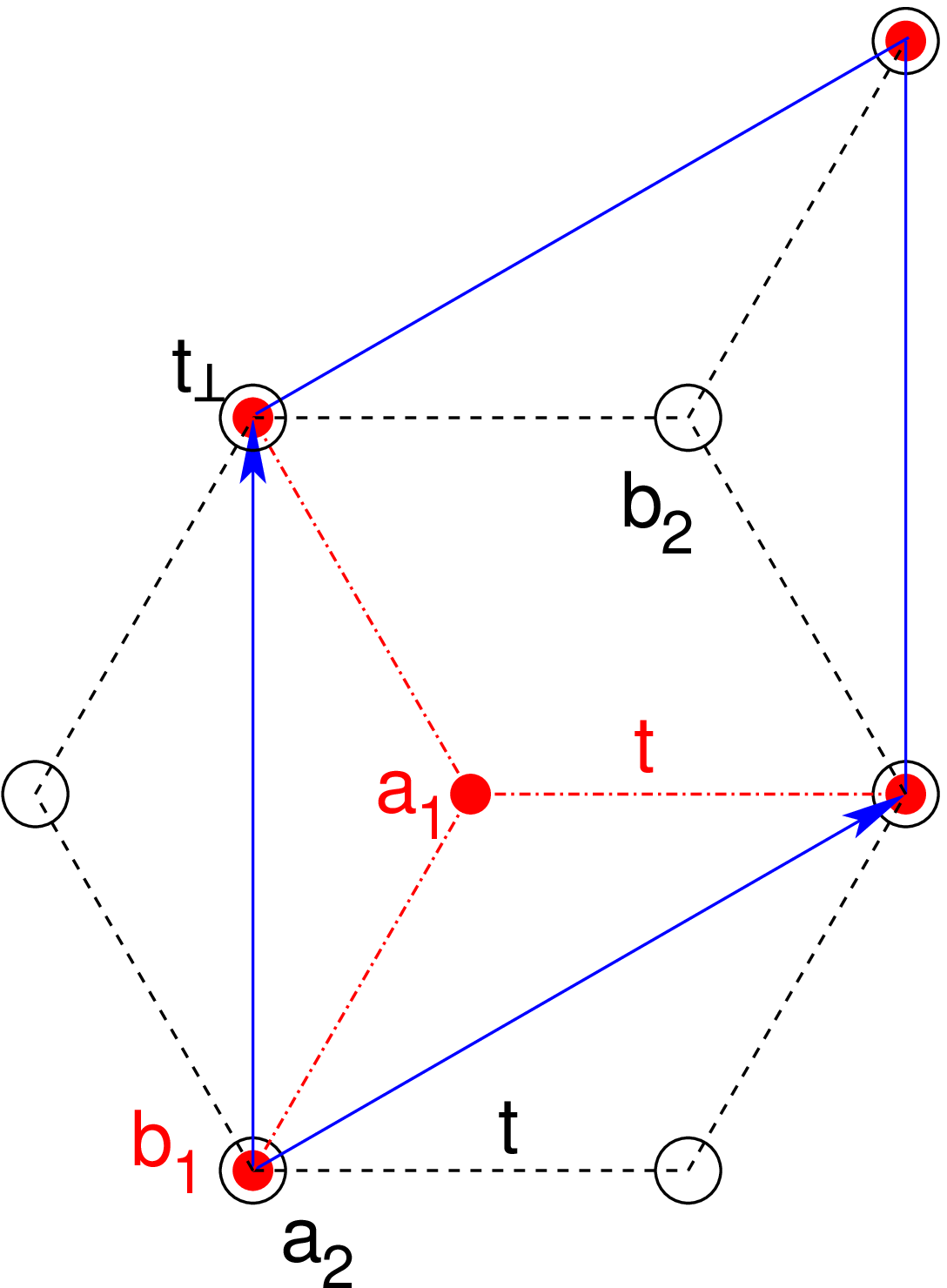}&
\includegraphics[width=0.3\textwidth]{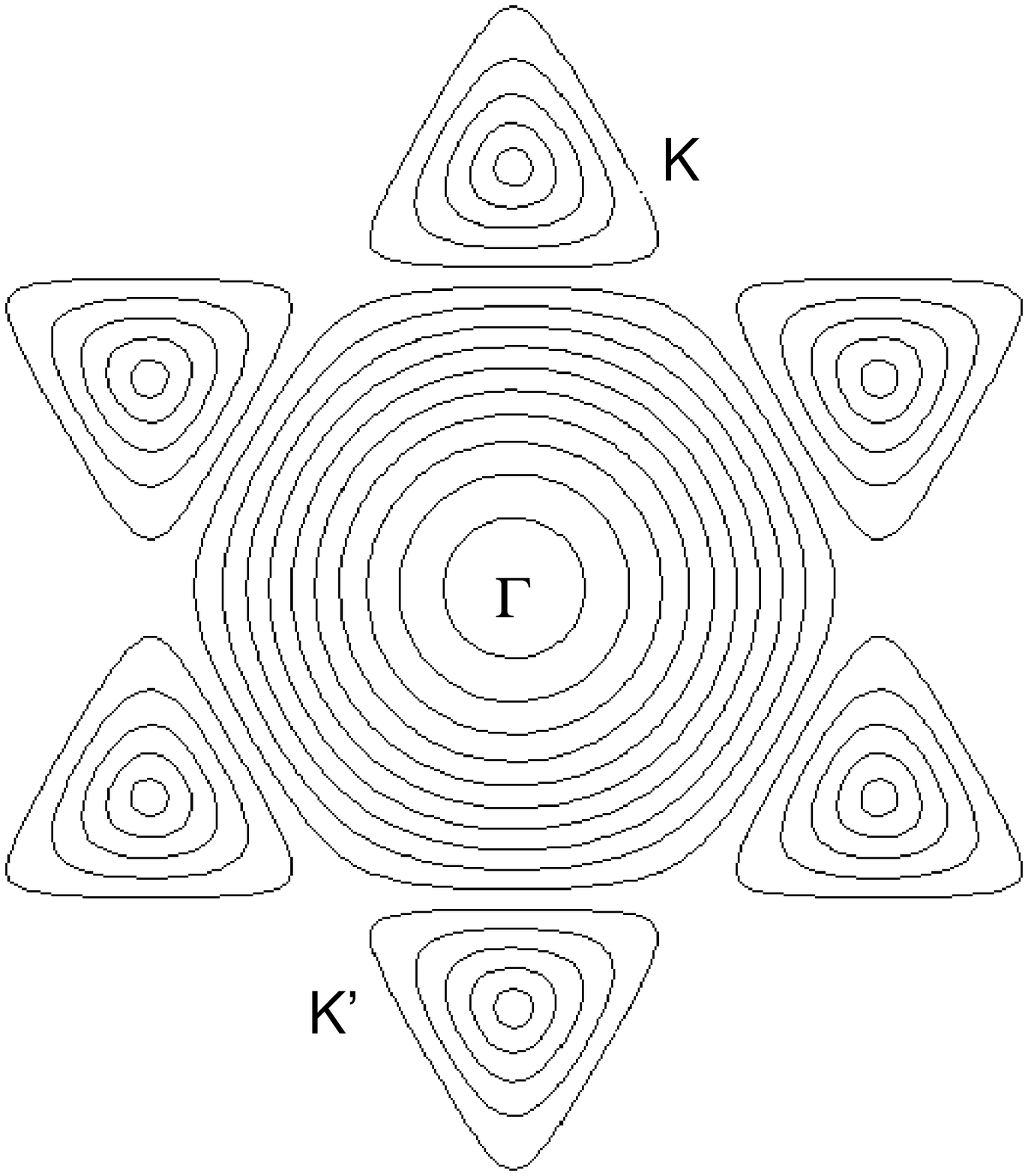}
\end{tabular}
\begin{tabular}{c}
\includegraphics[width=0.5\textwidth]{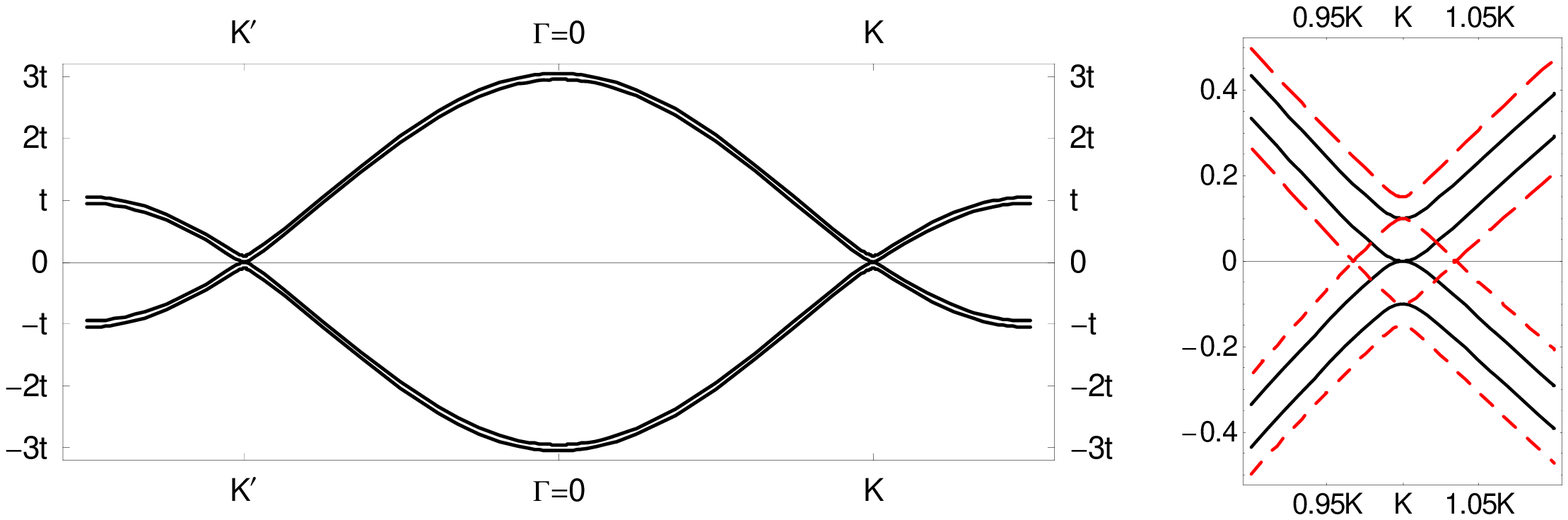}\end{tabular}
\end{center}
  \caption{(Upper left) Honeycomb bilayer unit cell. Atoms in the lower layer (2) are marked as
  empty (black) circles, atoms in the upper layer (1) are filled (red) circles. As a starting point,
  only the intralayer nearest neighbor hopping amplitudes $t$ and the interlayer
  hopping amplitudes $t_{\perp}$ are considered. (Upper right) Constant energy contours of the
  resulting dispersion, with minima at $K=\frac{4\pi}{\sqrt{3}3a}\hat{y}$ and $K'$ points and maximum at $\Gamma$ point.
  (Lower left) The energy dispersion of the four bands along the vertical cut in the Brillouin zone.
  The band splitting at the $K$ (and $K'$) points is $t_{\perp}$.
  (Lower right) Magnification of the dispersion (in units of $t$) near the degeneracy point (solid black) as well as the
  dispersion in the nematic state (dashed red) with $\Delta_x\neq 0$ (See Eq.\ref{eq:OpX}). }\label{unitcell}
\end{figure}

In this paper we apply the RG method to the bilayer graphene with
Bernal
stacking\cite{novoselovBilayerNatPhys2006,mccannBilayerPRL2006,netoRMP,GeimMacDonaldPhysToday07}.
While in general, the motion of the non-interacting electrons in
such potential does not lead to diverging susceptibilities since the
energy spectrum has two sets of four Dirac points in the corners of
the Brillouin zone (due to trigonal
warping)\cite{mccannBilayerPRL2006,netoRMP}, if only nearest
neighbor hopping is considered each set of four Dirac points merges
into a single degenerate point with parabolic dispersion (See Fig.
\ref{unitcell}). As the nearest neighbor hopping amplitudes are the
largest, the latter is the natural starting point of theoretical
analysis\cite{nilsson2006,minBilayer}.

We start with the tight-binding Hamiltonian for electrons hopping on
the bilayer honeycomb lattice with Bernal stacking
\begin{equation}\label{latticeHam}
\mathcal{H}=\sum_{\langle
\br\br'\rangle}\left[t_{\br\br'}c_{\sigma}^{\dagger}(\br)c_{\sigma}(\br')+h.c.
\right]+\frac{1}{2}\sum_{\br\br'}\delta
\hat{n}(\br)V(\br-\br')\delta \hat{n}(\br'),
\end{equation}
where, in the nearest neighbor approximation, the (real) hopping
amplitudes $t$ connect the in-plane nearest neighbor sites belonging
to different sublattices and, for one of the sublattices, also the
sites vertically above it with amplitude $t_{\perp}$. Since there
are four sites in the unit cell, there are four bands whose
dispersion for the above model comes from the solution of the
eigenvalue problem:
\begin{eqnarray}
\left[\begin{array}{cccc}
0 & d^*_{\bk} & t_{\perp} & 0\\
d_{\bk} & 0 & 0 & 0\\
t_{\perp} & 0 & 0 & d_{\bk}\\
0 & 0 & d^*_{\bk} & 0
\end{array}\right]
\left[\begin{array}{c}
b_1(\bk)\\
a_1(\bk)\\
a_2(\bk)\\
b_2(\bk)
\end{array}\right]=E(\bk)\left[\begin{array}{c}
b_1(\bk)\\
a_1(\bk)\\
a_2(\bk)\\
b_2(\bk)
\end{array}\right].
\end{eqnarray}
We find $E(\bk)=\pm \left(\frac{1}{2}t_{\perp}\pm
\sqrt{|d_{\bk}|^2+\frac{1}{4}t^2_{\perp}}\right)$, with
$d_{\bk}=t\left[2\cos\left(\frac{\sqrt{3}}{2}k_ya\right)e^{-\frac{i}{2}k_xa}+e^{ik_xa}\right]$.
Two of the bands are gapped (at $\bK,\bK'$ by $t_{\perp}$) and
become separated from the low energy pair which touches at $\bk=0$
(See Fig.\ref{unitcell}). The resulting density of states at zero
energy is therefore finite.

The repulsive interaction $V(\br-\br')$ in Eq.(\ref{latticeHam}) is
taken to have a finite range $\xi$ which is however much larger than
the lattice spacing $a$. This is assumed to be the correct starting
point, since the full Coulomb interactions is
screened\cite{hwangBilayerScreening} at low energy due to the finite
density of states. The analysis starting from the $1/|\br-\br'|$
interaction will be postponed to a future publication.

\begin{figure}[t]
\begin{center}
\begin{tabular}{ccccc}
\includegraphics[width=0.1\textwidth]{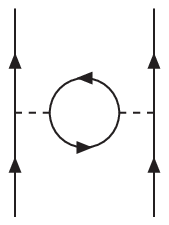}&
\includegraphics[width=0.09\textwidth]{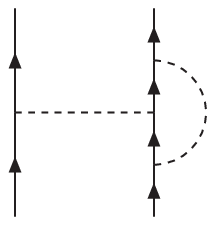}&
\includegraphics[width=0.09\textwidth]{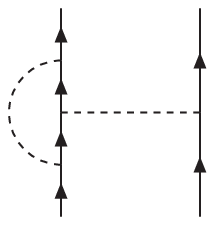}&
\includegraphics[width=0.07\textwidth]{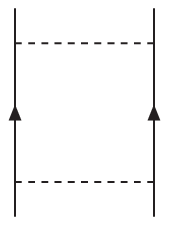}&
\includegraphics[width=0.07\textwidth]{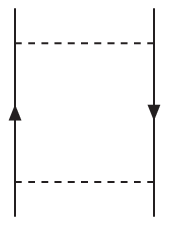}
\end{tabular}
\end{center}
  \caption{Diagrams appearing at 1-loop RG. The vertices are either
  $\delta_{\alpha\beta}$ or $\Sigma^{\mu}_{\alpha\beta}$.
  }\label{graphs}
\end{figure}
Following Nilsson {\it et. al.}\cite{nilsson:045405} we project out
the gapped bands. The resulting low energy effective (imaginary
time) action (which includes both $K$ and $K'$ valleys) is
\begin{eqnarray}
\mathcal{S}&=&\int d\tau d^2\br
\left[\psi^{\dagger}\left(\frac{\partial}{\partial
\tau}+\sum_{a=x,y}\Sigma^ad^a_{\bp}\psi\right)\right]\nonumber\\
&+&\frac{1}{2}g_1\int d\tau d^2\br
\psi^{\dagger}\psi(\br,\tau)\psi^{\dagger}\psi(\br,\tau)\nonumber\\
&+&\frac{1}{2}g_2\int d\tau d^2\br
\psi^{\dagger}\Sigma^z\psi(\br,\tau)\psi^{\dagger}\Sigma^z\psi(\br,\tau)\nonumber\\
&+&\frac{1}{2}g_3\int d\tau d^2\br
\sum_{a=x,y}\psi^{\dagger}\Sigma^a\psi(\br,\tau)\psi^{\dagger}\Sigma^a\psi(\br,\tau)
\end{eqnarray}
where the four component Fermi (Grassman) fields
\begin{eqnarray}
\psi(\br,\tau)=\int^{\Lambda}_0\frac{d^2\bk}{(2\pi)^2}e^{i\bk\cdot\br}
\left[\begin{array}{c}
a_1(\bK+\bk,\tau)\\
b_2(\bK+\bk,\tau)\\
a_1(\bK'+\bk,\tau)\\
b_2(\bK'+\bk,\tau)
\end{array}\right]
\end{eqnarray}
and
\begin{eqnarray}
d^x_{\bk}&=&\frac{k^2_x-k^2_y}{2m},\;\;\;
d^y_{\bk}=\frac{2k_xk_y}{2m},\\
\Sigma^x&=&1\sigma^x,\;\; \Sigma^y=\tau^z\sigma^y,\;\;
\Sigma^z=\tau^z\sigma^z.
\end{eqnarray}
The Pauli matrices $\sigma_j$ act on the layer indices $1$-$2$ and
the $\tau$ matrices act on the valley indices $\bK$-$\bK'$. The
effective mass is $m=2t_{\perp}/(9t^2)$, and $\psi$ represents
$\frac{N}{2}-$copies of the four component pseudo-spinor. $N=4$ for
spin $1/2$, and e.g. for $s=1,\ldots N$,
$\psi^{\dagger}\Sigma^z\psi(\br,\tau)=\psi_{\alpha
s}^{\dagger}\Sigma_{\alpha\beta}^z\psi_{\beta s}.$ Note that
$\Sigma's$ have the same multiplication table as the Pauli
$\sigma's$:
$\Sigma^{\mu}\Sigma^{\nu}=1_4\delta_{\mu\nu}+i\epsilon_{\mu\nu\lambda}\Sigma^{\lambda}$
and are traceless, too. $\Lambda$ is a momentum cutoff which
restricts the modes to the vicinity of the $\bK$-$\bK'$ points and
whose order of magnitude is $\lesssim \sqrt{2mt_{\perp}}$.

The coupling constant $g_{1}=\int d^2\br V(\br)$, i.e. it is the
$\bq=0$ Fourier component of $V(\br)$. The coupling constants $g_2$
and $g_3$ are zero in the starting action, but as will be shown
next, they get generated in the momentum-shell
RG\cite{Shankar.RevModPhys.66.129.1994}, and therefore they are made
explicit in the original action.

From simple power-counting, the (engineering) scaling dimension of
the field $\psi$ is $L^{-1}$ and $L^{2}$ for $\tau$. This makes
$g_1$, $g_2$ and $g_3$ marginal (at the tree-level) and the question
is how they flow upon inclusion of the loop corrections. To answer
this we note that all possible Wick
contractions\cite{Shankar.RevModPhys.66.129.1994} of four-fermion
operators correspond to the diagrams in the Figure (\ref{graphs}).
The RG equations obtained by integrating fermion modes within a thin
shell $\Lambda$ and $\Lambda/s$ (centered at the $K$ point), and
$\int^{\infty}_{-\infty}\frac{d\omega}{2\pi}$, are:
\begin{eqnarray}
\frac{dg_1}{d\ln s}&=&\left[-4g_1g_3\right]\frac{m}{4\pi}\label{RGs_1}\\
\frac{dg_2}{d\ln s}&=&\left[-4(N-1)g^2_2+4g^2_3+4g_1g_2-12g_2g_3\right]\frac{m}{4\pi}\label{RGs_2}\\
\frac{dg_3}{d\ln
s}&=&\left[-(g_1-g_3)^2-(g_2-g_3)^2-2(N+1)g^2_3\right]\frac{m}{4\pi}\label{RGs_3}
\end{eqnarray}
\begin{figure}[t]
\begin{center}
\includegraphics[width=0.5\textwidth]{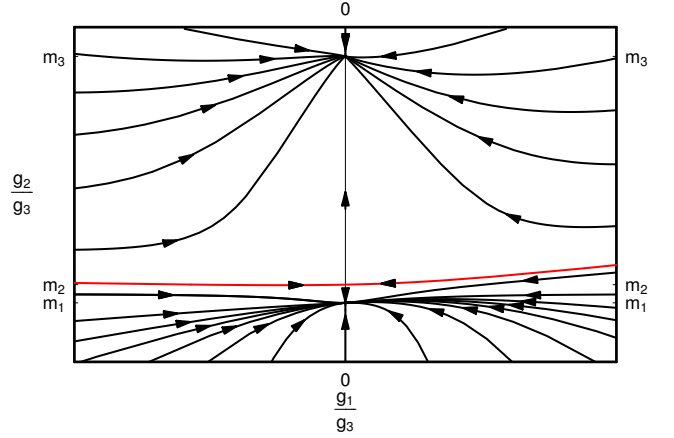}
\end{center}
  \caption{RG flow diagram of the ratios $g_1/g_3$ and $g_2/g_3$ for $g_3<0$. While the ratio $g_1/g_3$ flows to zero
  (even if the starting point is $g_2=g_3=0$ and $g_1\neq 0$), the ratio $g_2/g_3$ flows to a fixed value, indicating two stable and one unstable rays
  with slopes $m_1\approx -0.525$, $m_3\approx13.98$ and $m_2\approx0.545$, respectively.}\label{g1g2Flow}
\end{figure}

While the above equations cannot be solved in a closed form, it is
possible to fully analyze the qualitative nature of the RG flows.
Such analysis is facilitated by the observation that
$$\frac{dg_3}{d\ln s}\leq 0$$ which means that, unless $g_1=g_2=g_3=0$
when the equality holds, $g_3$ strictly decreases under RG
rescaling. We can therefore trade the parametric dependence on $s$
of $g_1$ and $g_2$ for their dependence on $g_3$ {\it and} retain
the direction of the RG flow. For $g_3<0$ ($>0$), an increase in
$d\log s$ therefore corresponds to an increase (decrease) in
$\frac{dg_3}{g_3}$. Since the system is autonomous, we can eliminate
$\log s$ and arrive at a system
\begin{eqnarray}\label{RGg3_1}
\frac{dg_1}{dg_3}=f\left(\frac{g_1}{g_3},\frac{g_2}{g_3}\right)\\
\frac{dg_2}{dg_3}=g\left(\frac{g_1}{g_3},\frac{g_2}{g_3}\right)\label{RGg3_2}
\end{eqnarray}
where
\begin{eqnarray}
f\left(x,y\right)&=&\frac{-4x}{-x^2-y^2-2(N+2)+2x+2y}\\
g\left(x,y\right)&=&\frac{-4(N-1)y^2+4+4xy-12y}{-x^2-y^2-2(N+2)+2x+2y}
\end{eqnarray}

The system of Eqs.(\ref{RGg3_1})-(\ref{RGg3_2}) is in turn
homogeneous and can therefore be written as
\begin{eqnarray}
g_3\frac{d\frac{g_1}{g_3}}{dg_3}=-\frac{g_1}{g_3}+f\left(\frac{g_1}{g_3},\frac{g_2}{g_3}\right)\\
g_3\frac{d\frac{g_2}{g_3}}{dg_3}=-\frac{g_2}{g_3}+g\left(\frac{g_1}{g_3},\frac{g_2}{g_3}\right).
\end{eqnarray}
The above system has three fixed points, all of which have
$g_1/g_3=0$, while $g_2/g_3=m_1,m_2,m_3$. As shown in the
Fig.(\ref{g1g2Flow}), $m_1\approx -0.525$ and $m_3\approx13.98$ are
sinks, while $m_2\approx0.545$ has one attractive direction and one
repulsive. This means that once $g_3$ gets to be negative, only
$g_2$ and $g_3$ become important (their ratio being fixed) while
$g_1$ is too small compared to $g_3$. To see that this is indeed
what happens if the starting point is $g_1(s=1)>0$ and
$g_2(s=1)=g_3(s=1)=0$, note that the Eqs.(\ref{RGs_1}-\ref{RGs_3})
imply that finite $g_1$ generates finite and negative $g_3$ upon
first iteration while $g_2$ remains zero until the second iteration.
This means that we start with $g_1/g_3\rightarrow -\infty$ and
$g_2/g_3=0$ which is {\it below} the (red) separatrix, thus the flow
is into the region of attraction of $m_1$ (Fig.(\ref{g1g2Flow})).

\begin{table}[t]
\begin{center}
    \begin{tabular}{ | c || c | c | c | c |}
    \hline
    $\psi^{\dagger}\tau^{\mu}\sigma^{\nu}\psi$ & $\nu=0$ & $\nu=x$ & $\nu=y$ & $\nu=z$\\
    \hline\hline
    $\mu=0$ & $0,0,0$  & $1,-1,-2N$ & $1,-1,0$  & $2,2,-4$\\ \hline
    $\mu=x$ & $1,-1,0$ & $0,0,0$    & $2,2,-4$  & $1,-1,0$ \\ \hline
    $\mu=y$ & $1,-1,0$ & $0,0,0$    & $2,2,-4$  & $1,-1,0$ \\ \hline
    $\mu=z$ & $0,0,0$  & $1,-1,0$   & $1,-1,-2N$& $2,2-4N,-4$\\\hline
    \end{tabular}
\end{center}
\caption{The susceptibility coefficients $A,B,C$ in
Eq.(\ref{eq:vertexGeneral_ph}) for different particle-hole order
parameters $\psi^{\dagger}\mathcal{O}_i\psi$. In the physical case
$N=4$.} \label{tab:ABC}
\end{table}
\begin{table}[t]
\begin{center}
    \begin{tabular}{ | c || c | c | c | c |}
    \hline
    $\psi_{\alpha s}(\tau^{\mu}\sigma^{\nu})_{\alpha\beta}\psi_{\beta s'}$ & $\nu=0$ & $\nu=x$ & $\nu=y$ & $\nu=z$\\
    \hline\hline
    $\mu=0$ & $-1,-1,0$  & $-2,2,-4$ & $0,0,0$   & $-1,-1,0$\\ \hline
    $\mu=x$ & $-2,2,-4$  & $-1,-1,0$ & $-1,-1,0$ & $0,0,0$ \\ \hline
    $\mu=y$ & $-2,2,-4$  & $-1,-1,0$ & $-1,-1,0$ & $0,0,0$ \\ \hline
    $\mu=z$ & $-1,-1,0$  & $-2,2,-4$ & $0,0,0$   & $-1,-1,0$\\\hline
    \end{tabular}
\end{center}
\caption{The susceptibility coefficients $A',B',C'$ in
Eq.(\ref{eq:vertexGeneral_pp}) for different particle-particle order
parameters
$\psi_{\alpha\sigma}\mathcal{O}^{(i)}_{\alpha\beta}\psi_{\beta\sigma'}$.
} \label{tab:ApBpCp}
\end{table}
\begin{figure}[t]
\begin{center}
\includegraphics[width=0.43\textwidth]{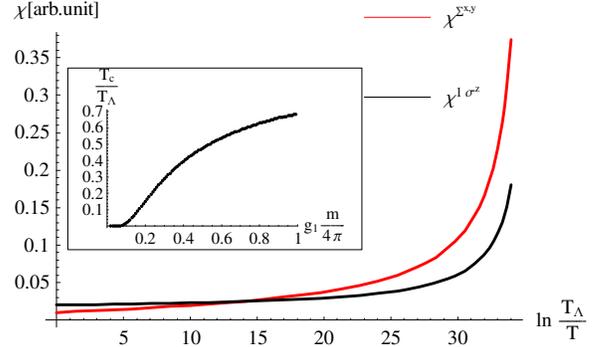}
\end{center}
  \caption{Numerical integration of the susceptibilities in Eq.(\ref{eq:vertexGeneral_ph})
  for $g_1(s=1)=0.01$ and $g_2(s=1)=g_3(s=1)=0$.
The strongest divergence is towards the nematic order. (Inset)
Numerically determined nematic transition temperature in units of
cutoff $T_\Lambda\lesssim t_{\perp}$ as a function of the
dimensionless coupling $g_1\frac{m}{4\pi}$.
  }\label{fig:vertices}
\end{figure}

From Eqs.(\ref{RGs_1}-\ref{RGs_3}) we see for the fixed ratios
$g_1/g_3=0$ and $g_2/g_3=m_j$, $g_3$ becomes large and negative,
indicating a runaway flow. Given the flow of the coupling constants
we can determine the susceptibilities towards the formation of
ordered states. In particular, we consider coupling the fermions to
external sources, which correspond to the possible broken symmetry
states. We therefore have additional terms in the action:
\begin{eqnarray}
\Delta{\mathcal S}&=&-\Delta_{ph}^{\mathcal{O}_i}\int d\tau d^2\br
\psi^{\dagger}\mathcal{O}_i\psi(\br,\tau)\nonumber\\
&-&\Delta^{\mathcal{O}_i}_{pp}\int d\tau d^2\br
\psi_{\alpha\sigma}\mathcal{O}^i_{\alpha\beta}\psi_{\beta\sigma'}(\br,\tau)
\end{eqnarray}
Such terms, with infinitesimal $\Delta$'s explicitly break the
symmetry and so are relevant operators. The question of instability
is answered by finding the renormalization of the
vertices\cite{chubukov-2009}. The one which diverges first
determines the broken symmetry states. After a straigthforward
calculation we find that
for a general particle-hole order parameter
$\mathcal{O}_i=\tau^{\mu}\sigma^{\nu}$ where $\mu,\nu=0,1,2,3$ and
$\tau_0=\sigma_0=1$,
\begin{eqnarray}\label{eq:vertex1}
\Delta_{ph,ren}^{\tau^{\mu}\sigma^{\nu}}&=&\Delta_{ph}^{\tau^{\mu}\sigma^{\nu}}\left(1+\left[Ag_1+Bg_2+Cg_3\right]\frac{m}{4\pi}\ln
s\right)\label{eq:vertexGeneral_ph}\nonumber\\
\end{eqnarray}
where the coefficients $A$, $B$, and $C$ are given in the Table
\ref{tab:ABC}. Similarly, for a general particle-particle order
parameter
$\psi_{\alpha\sigma}\mathcal{O}^{(i)}_{\alpha\beta}\psi_{\beta\sigma'}$
\begin{eqnarray}\label{eq:vertex1}
\Delta_{pp,ren}^{\tau^{\mu}\sigma^{\nu}}&=&\Delta_{pp}^{\tau^{\mu}\sigma^{\nu}}\left(1+\left[A'g_1+B'g_2+C'g_3\right]\frac{m}{4\pi}\ln
s\right)\label{eq:vertexGeneral_pp}\nonumber\\
\end{eqnarray}
where the coefficients $A'$, $B'$, and $C'$ are given in the Table
\ref{tab:ApBpCp}.

The instability towards a particular order occurs at an energy scale
({\it i.e.} temperature) at which the corresponding coefficient of
the $\ln s$ in
Eqs.(\ref{eq:vertexGeneral_ph}-\ref{eq:vertexGeneral_pp}) diverges.
Since $N=4$ and the fixed point value of $g_2/g_3\approx -0.525$,
with $g_3$ large and negative, it can be seen from Table
\ref{tab:ABC} that the instability appears in the $\Sigma^{x,y}$
channel, which as we discuss next corresponds to a {\it nematic}
order. The numerical integration of the RG equations
(\ref{RGs_1}-\ref{RGs_3}) starting with $g_1(s=1)>0$ and
$g_2(s=1)=g_3(s=1)=0$ shown in Fig.(\ref{fig:vertices}) indeed
confirms that the susceptibility diverges fastest in this channel.
Within the continuum model and in weak coupling, the instability is
therefore towards the order parameter, which we can parametrize by a
complex field
$$\Delta_{nem}(\br)\equiv\Delta_x(\br)+i\Delta_y(\br)=\langle
\psi^{\dagger}(\br)\left(\Sigma^x+i\Sigma^y\right)\psi(\br)\rangle.$$
To see that this is indeed a nematic order, note that at $\bq=0$ (1)
it is translationally invariant and (2) even under rotations by
$\pi$. In fact, as the low energy Hamitonian is invariant under
arbitrary rotations by an angle $\alpha$, i.e. $
U^{\dagger}(\alpha)\mathcal{H}U(\alpha)=\mathcal{H}$, where $
U_{\alpha}=e^{-i\alpha
\hat{L}_z}e^{-i\alpha\Sigma^z},\;\;L_z=x\frac{\partial}{\partial
y}-y\frac{\partial}{\partial x}$, we find that under a rotation by
$\alpha$
$$\Delta_{nem}(\br)\rightarrow \Delta_{nem}(\br)e^{2i\alpha}.
$$
This shows that the order parameter is even under rotations by $\pi$
and odd under rotations by $\pi/2$, which makes it nematic. For
uniform $\Delta_{nem}(\br)$ the quadratic degeneracy point is split
into two (massless) Dirac points by an amount proportional to the
magnitude of the order parameter and the direction given by the
nematic director.

The presence of the underlaying lattice further breaks the full
rotational symmetry of the long distance effective Hamiltonian down
to hexagonal symmetry centered on $a_2-b_1$ site, where the standard
operations of $C_{6v}$ must be accompanied by the appropriate layer
permutations. The two components of the order parameter, which give
finite expectation values of, for instance, $\Delta_x(\br)=$
\begin{eqnarray}
&&\left\langle a^{\dagger}_{1\sigma}(\br)\left(b_{2\sigma}(\br-a\hat{x})-\frac{1}{2}\sum_{s=\pm}b_{2\sigma}(\br+\frac{a}{2}\hat{x}s\frac{\sqrt{3}}{2}\hat{y})\right)
+h.c.\right\rangle\nonumber\\
\label{eq:OpX}\\
&&\mbox{and}\;\; \Delta_y(\br)=\nonumber\\
&&\left\langle
a^{\dagger}_{1\sigma}(\br)\left(\frac{\sqrt{3}}{2}\sum_{s=\pm}sb_{2\sigma}(\br+\frac{a}{2}\hat{x}
+s\frac{\sqrt{3}}{2}\hat{y})\right)+h.c.\right\rangle\label{eq:OpY}
\end{eqnarray}
form a two dimensional representation of the hexagonal group. Note
that the nematic order parameter remains even under $\pi$-rotation
followed by the layer permutation.

From the arguments above we expect that the lattice has an important
effect on the critical nature of the phase transition, which would
otherwise be of Kosterlitz-Thousless kind. The reason is the
existence of the third order invariant
$\Delta^3_x-3\Delta_x\Delta^2_y$. As a result the finite temperature
phase transition should be described by the effective Hamiltonian
\begin{eqnarray}
\mathcal{H}_{nem}=\sum_{\langle \bx\bx'\rangle}
-J\cos[2(\theta(\bx)-\theta(\bx'))]+h\sum_{\bx}\cos[6\theta(\bx)].
\end{eqnarray}
where $\Delta_x(\bx)+i\Delta_y(\bx)=e^{2i\theta(\bx)}$,
$\theta\in(0,2\pi]$ and the sum runs over the vertices of the
triangular sub-lattice spanned by $a_1$ sites. This corresponds to
the $p=3$ case of the two dimensional planar model studied by Jose
et.al.\cite{JosePRB1977} and the concomitant absence of the Gaussian
spin-wave phase. Instead there is a continuous transition between
the low temperature phase where the director locks into one of three
values and a high temperature phase where vortices unbind. Such
transition is believed to belong to the 2D three-state Potts model
universality class\cite{NelsonBook} with
exponents\cite{Wu_RMP_Potts1982} $\nu=5/6$ and $\eta=4/15$.

Finally, we discuss the effects of the trigonal warping which splits
each of the quadratic degeneracies into four massless Dirac points,
which were ignored up to now. If we denote the energy scale
associated with such terms as $T_{trig}$, below which the dispersion
must be modified, then the transition will still occur provided that
the mean-field transition temperature $T_c$ estimated from the above
model and plotted in the inset of Fig.(\ref{fig:vertices}) satisfies
$T_c\gg T_{trig}$. For screened Coulomb
interactions\cite{hwangBilayerScreening} $g_1\frac{m}{4\pi}\sim
\mathcal{O}(1)$, leading to $T_c\lesssim t_{\perp}$. Since the
current estimates of $T_{trig}$ are of the same order of
magnitude\cite{zhang:235408}, the ultimate test is experimental.

Acknowledgements: While this paper was in preparation, we became
aware of Ref.\cite{sun-2009} where lattices with fourfold and
sixfold rotational symmetry are constructed in either case the
parabolic degeneracy points are protected by the point group
symmetry. In there, the degeneracy point maps unto itself under time
reversal, unlike our $K$ and $K'$, and nematic was found to be
stabilized (within mean-field) only at finite coupling. This work is
supported in part by NSF grant No. DMR-0704133 (KY). Part of this
work was carried out while the authors were visiting Kavli Institute
for Theoretical Physics (KITP). The work at KITP is supported in
part by NSF grant No. PHY-0551164.
\bibliography{bilayer}
\end{document}